\documentclass{article}
\usepackage{graphicx}
\usepackage{epsfig}
\usepackage[usenames]{color}
\usepackage{amsmath, amssymb}
\usepackage{fancyvrb}
\usepackage{cite}
\usepackage{url}

\newcommand{\cc}{{\mathbb C}}

\newcommand{\bb}{{\bf b}}
\newcommand{\bx}{{\bf x}}

\DeclareFontShape{OT1}{cmtt}{bx}{n}{<5><6><7><8><9><10><10.95><12><14.4><17.28><20.74><24.88>cmttb10}{}

\newcommand{\ttbf}[1]{\texttt{\textbf{#1}}}

\newcounter{figctr}
\newcommand{\ilab}[1]{\immediate\write1{\string \newlabel{#1}{{\arabic{figctr}}{\thepage}}}}

\newcommand{\chhobi}[3]{
1\parbox{\textwidth}{\begin{center}
  \scalebox{#2}{\includegraphics{#1}}\\
\underline{{\bf Fig \arabic{figctr}:\ \ilab{im:#1}\addtocounter{figctr}{1}}
  {\bf #3}}
\end{center}}
}

\title{QuECT: A New Quantum Programming Paradigm}

\author{Arnab Chakraborty\\
{\em Applied Statistics Unit}\\
{\em Indian Statistical Institute}\\
{\em Kolkata, India 700108}\\
{\tt arnabc@isical.ac.in}}

\begin{document}

\maketitle

\abstract{Quantum computation constitutes a rapidly expanding
  subfield of computer science. Development quantum
  algorithms is facilitated by the availability of efficient
  quantum programming languages, and a plethora of approaches
  has been already suggested in the literature, ranging from
  GUI-based simple tools to elaborate standalone programming
  languages. In this paper we propose a novel paradigm called
  Quantum Embeddable Circuit Technique (QuECT) that allows a
  programmer to embed a circuit diagram in a classical ``host''
  language. The paradigm can be implemented in any modern
  classical language. A prototype has been developed by the
  author using Java.
\section{Introduction}

Quantum computation is a rapidly developing field, and even
though actual quantum computers are yet to be constructed in a
large scale, many quantum computing algorithms have already appeared in the
literature. One needs some quantum programming language to
present these algorithms to a quantum computer (or a
simulator). A variety of approaches have been suggested in the
literature to this end. These range from drag-and-drop graphical
interfaces to quantum assembly languages.  Most textbooks
like to present quantum algorithms in the form of ``circuit
diagrams.'' All the graphical approaches to quantum programming
rely on these circuit diagrams.
Its intuitive nature notwithstanding, this ubiquitous technique is not  easily amenable to
standard programming constructs like conditional jumps and
loops. 
In this paper we propose a new quantum programming paradigm
called {\bf Quantum Embeddable Circuit Technique} {\bf (QuECT)}
that combines the power of circuit diagram and yet allows the
traditional control structures. It  enhances  an existing classical
language  to handle quantum circuit diagram embedded in it.
 The author's implementation uses
Java as the classical ``host'' language, but the same idea carries
over directly to any other language (compiled/byte code
compiled/interpreted). The QuECT paradigm is equally applicable
for a real quantum computer (when one is
constructed) as it is for a simulator. It is expected that this
paradigm will help traditional programmers to pick up
quantum computing more easily.

At this point the reader may like to take a look at a simple
example of the QuECT approach given in section 4.
The layout of the paper is
as follows. Section 2 gives a concise summary of the basic
notions of quantum computing. Our exposition loosely follows that
of \cite{noson}. In section 3 we review the quantum programming
paradigms in existence.
The section after that is devoted to the details of
the QuECT paradigm. Section 5 shows the paradigm in action, where
we demonstrate QuECT implementation of some standard quantum
algorithms. Some practical issues about compiling a QuECT program
may be found section 7.

\section{Basic ideas}
\subsection{Qubits}
The most fundamental building block of quantum computing  is
a {\bf qubit}, which is often considered as the quantum analog
of a bit, and hence its name. This analogy, however, has its
pitfalls, and for a rigorous understanding one should better
remember a qubit as a {\em nonzero} element of $\cc^2,$
identified up to multiples. Thus, $(1,i)$ is a qubit,
and $(1,2+i)$ is another. The qubit $(i,-1)$ is
actually the same as $(1,i),$ since these are multiples of
each other. 
The two qubits $(1,0)$ and $(0,1)$ are special, and
have the names $|0\rangle$ and $|1\rangle,$ respectively.

By a $k$-qubit register we shall understand a {\em nonzero}
element of $\cc^{2^k}.$ A $1$-qubit register is the
same as a single qubit. However, for general $k$
a $k$-qubit register is {\em not} the same as a
collection of $k$ qubits. Two $k$-qubit registers are
considered the same if they are (nonzero) multiples of each
other. 

Since we are working with vectors and matrices of size $2^k,$ it is
obvious that the sizes are going to explode pretty soon even for
a modest $k.$ While it is not supposed to pose a problem to
a computer with quantum hardware, it is
nevertheless difficult for a human programmer to keep track of such
large vectors and matrices. The concept of factorization helps us
here by expressing a large vector as {\em tensor product} (also known
as {\em Kronecker product}) of shorter vectors. This
allows us to work with the shorter vectors separately,
combining the results at the very end, if necessary. 
For example, it may be possible to express a $n$-qubit register (consisting of $2^n$
complex numbers) may be expressed as a tensor product of $k$
registers of sizes $n_1,...,n_k$ (where $\sum n_i = n)$ 
requiring a total of only $\sum 2^{n_i}$ complex numbers).

However, not all quantum registers admit a factorization. If a $n$-qubit register
  cannot be factored, then we say that the $n$ qubits
  are {\bf entangled}. Entangled qubits give the main power to
  quantum programming, and also stand as the main hurdle for
  a classical programmer aspiring to write a
  quantum program.

\subsection{Quantum gates}
The second most important concept in quantum computing is that of
quantum gates, which are unitary matrices with complex
entries. We shall think of a $k$-qubit register as a column
vector of length $2^k,$ and a quantum gate as a unitary
matrix of size $2^k$ multiplying it from the left. From the
viewpoint of a quantum programmer these are of two types. First,
there are some commonly used operators ({\em e.g.}, Hadamard,
Pauli's $X,Y,Z$ etc). Then there are the operators  specially
crafted for some specific algorithm. The most prominent example
of the later type is a ``quantum wrapper'' that packages
a classical function $f:\{0,1\}^m\to \{0,1\}^n$ into 
a unitary
matrix $U_f$ of order $2^{m+n}$ as follows. 

To find
the $(i,j)$-th entry of $U_f$ we first express $i$
and $j$ in binary using $m+n$ bits. Then we split each
of them into two parts: the $x$-part consisting of the most
significant $m$ bits, and the $y$-part consisting of
the least significant $n$ bits. Let these be denoted by 
$i_x,i_y,j_x$ and $j_y.$ Then 
$$
U(f)_{ij} = \left\{\begin{array}{ll}
1 &\mbox{if }i_x=j_x \text{ and } f(i_x)=i_y\oplus j_y\\
0 &\mbox{otherwise.}
\end{array}\right.
$$
It is not difficult to see that each row and each column
of $U(f)$ has exactly a single 1, and so $U(f)$ is
unitary. 

\subsection{Measurement}
The third important concept in quantum computing is that of
measurement. While it is standard to treat measurement as a
Hermitian operator, we shall restrict ourselves to the most
frequently used form, {\em viz.}, measurement of one or more qubits in a
multi-qubit register. If we measure $p$ given qubits in
a $k$-qubit register then we shall observe a random variable
which takes values $0,1,...,2^p-1$ (or, equivalently, the
corresponding bit patterns of length $p$). Quantum physics
dictates that the probability
of observing a given bit pattern $\bb$ is 
$$
\frac{\sum' |z_i|^2}{\|z\|^2},
$$
where the numerator sum is for those $i$'s only whose binary
representations have the pattern $b$ in the positions of the
qubits being measured. For example, if we measure qubits at
positions 0
and 2 in a 3-qubit register containing the value 
$(z_0,...,z_7),$ we shall observe 0 or 1 or 2 or 3 with the
probabilities
\newcommand{\ul}{\underline}
\begin{eqnarray*}
P(0) = P(00) &=& \frac{|z_{{\ul0}0{\ul0}}|^2+|z_{{\ul0}1{\ul0}}|^2}{\|z\|^2},\\
P(1) = P(01) &=& \frac{|z_{{\ul0}0{\ul1}}|^2+|z_{{\ul0}1{\ul1}}|^2}{\|z\|^2},\\
P(2) = P(10) &=& \frac{|z_{{\ul1}0{\ul0}}|^2+|z_{{\ul1}1{\ul0}}|^2}{\|z\|^2},\\
P(3) = P(11) &=& \frac{|z_{{\ul1}0{\ul1}}|^2+|z_{{\ul1}1{\ul1}}|^2}{\|z\|^2}.
\end{eqnarray*}
Here $z_{000}$ means $z_0,$ $z_{001}$
means $z_1,$ and so on. Also the bits at positions 0 and 2
are underlined in the right hand sides for ease of comparison. 
The process of measurment also changes the contents of the
multi-qubit register. If the output is the bit pattern $\bb,$
then the contents of the
multi-qubit register changes from $z$ to $w$ where
$$
w_i = \left\{\begin{array}{ll}
z_i &\mbox{if }i \text{ has pattern } \bb \text{ in the measured
      positions} \\
0 &\mbox{otherwise.}
\end{array}\right.
$$

The fact that a quantum measurement potentially changes the
underlying quantum register is a main distinguishing feature of quantum systems
over classical ones.
However, in a typical quantum algorithm the measurement comes
at the very end. So we do not care about the fate of the
register once the measurement is over. However the way a
measurement affects the contents of a quantum register has the
following theoretical implication which indirectly helps a quantum
programmer.

Suppose that we are working with a $k$-qubit
register. Let $A,B$ be two disjoint, nonempty subsets
of $\{0,...,k-1\}.$ Then all the following measurement
operations will produce identically distributed outputs:

\begin{enumerate}

\item First measure the qubits at positions $A,$ then measure
  those at positions $B.$
\item Measure the qubits at positions $A\cup B.$

\end{enumerate}

This observation allows us to combine all the measurements at the
end of an algorithm into a
single mesaurement.
\subsection{Quantum circuits}
A quantum algorithm, in its barest form, consists of 

\begin{enumerate}

\item a multi-qubit register with some given classical initial
  value, ({\em i.e.}, a tensor product of $|0\rangle$'s and $|1\rangle$'s),

\item an ordered list of quantum gates that act on the register in
  that order,
\item measurements of some qubits at the very end.

\end{enumerate}

Usually the quantum gates are complicated, huge matrices, made by taking
tensor products of smaller matrices. An actual quantum computer
will have no problem in computing the effect of applying such
gates, but a human programmer always finds it easier to specify
a huge gate in terms of the constituent smaller gates. The
quantum circuit diagram is the most popular way to achieve this. 

The evolution of a $k$-qubit
register is shown in a quantum circuit as $k$ parallel
lines. Roughly speaking, these lines are like $k$ wires
each carrying a single qubit\footnote{This analogy is not entirely correct
in presence of possible entanglement, as we shall see soon.}.
 The left hand circuit diagram in {\rm Fig \ref{im:ckt1.eps}}, for example, represents
a 1-qubit register initially storing $|0\rangle,$ and being
acted upon by a quantum gate $A$ 
(which is a $2\times2$ unitary matrix). 

\chhobi{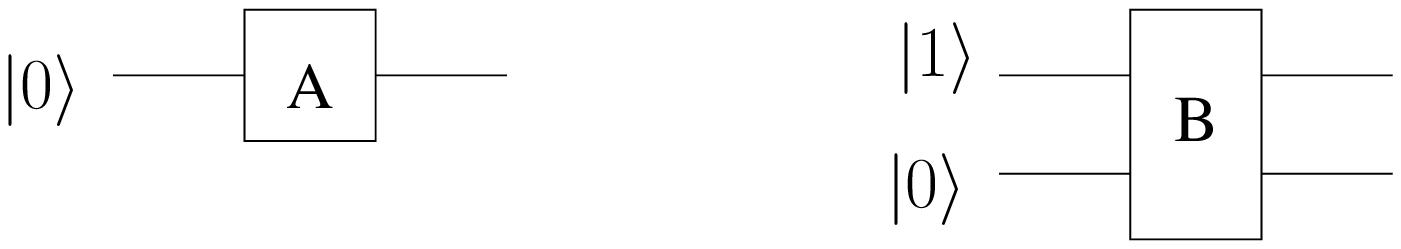}
       {0.4}
       {Two simple quantum
  circuits}
 
The right hand circuit diagram represents a 2-qubit register
starting with the value 
$$
|1\rangle\otimes|0\rangle = \left[\begin{array}{cccccc}
0\\1\\0\\0
\end{array}\right]
$$
and being acted upon by a quantum gate $B$ which must be
a $2^2\times 2^2$ unitary matrix. The result may not be
factorizable as a tensor product of two qubits, in which case it
is meaningless to talk about the values of the individual output
lines. 

The following circuit shows a 3-qubit register starting with the
content 
$$
|0\rangle\otimes|1\rangle\otimes|0\rangle = \left[\begin{array}{cccccc}
0\\1\\0\\0\\0\\0\\0\\0
\end{array}\right].
$$

\chhobi{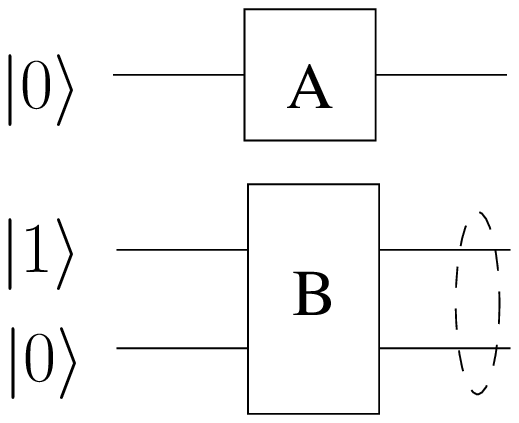}
       {0.4}
       {An example of entanglement}

Then it is acted upon by the $2^3\times 2^3$ unitary matrix
$$
A\otimes B.
$$

The lower 2 qubits are possibly entangled together, however they
are not entangled with the top qubit. This is shown using a
broken ellipse (which is not part of a standard circuit
diagram). 

{\rm Fig \ref{im:ckt3.eps}} presents a more complicated example that also
demonstrates a pitfall.

\chhobi{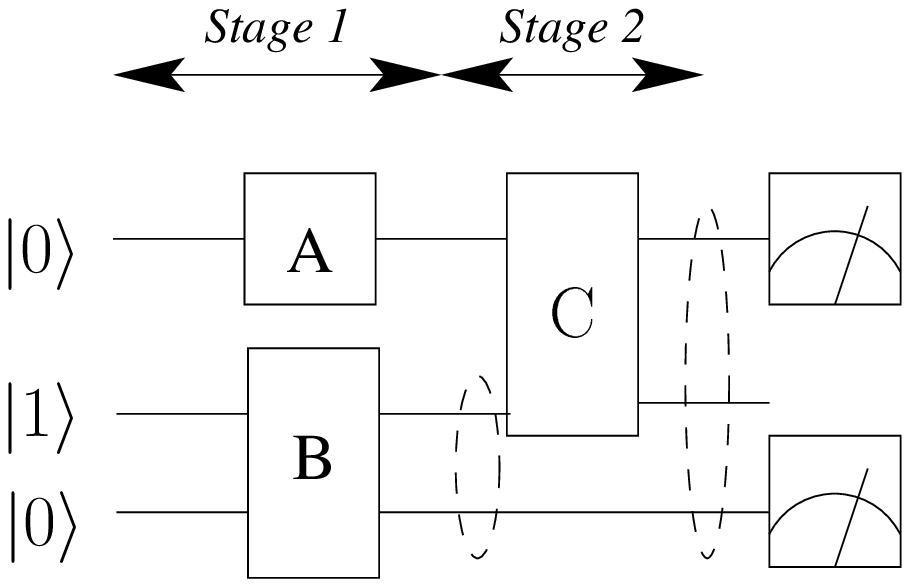}
       {0.4}
       {A more complicated circuit}

It would be wrong to think that the quantum gate $C$
acts upon the
top two qubits. This is because the middle qubit actually does
not exist separately after stage 1, thanks to the possible
entanglement introduced by $B.$ Thus the naive
interpretation that each line is like a wire carrying a single
qubit does not hold any more. The unitary matrix that is depicted
by this diagram is actually
\begin{equation*}\tag{{\bf *}}\label{*}
(C\otimes I_2)(A\otimes B),
\end{equation*} 
where $I_2$ denotes identity matrix of order 2
(corresponding to the bottom line ``passing through'' stage 2). This circuit
also has some measurement symbols. Here we are measuring the two
extreme qubits. The overall circuit denotes a quantum function
from $\{0,1\}^3$ to $\{0,1\}^2.$

The usefulness of circuit diagrams to represent quantum
algorithms stems from the following reasons.

\begin{enumerate}

\item It allows one to work with just $k$ lines even when the
  underlying system has dimension $2^k.$
\item The quantum gates can be constructed easily out of component
  gates suppressing cumbersome expressions like \eqref{*}
  involving tensor products and identity matrices.

\end{enumerate}

The chief drawbacks of a circuit diagram are as follows. 

\begin{enumerate}

\item Before talking about the contents of a subset of the lines
  one has to make sure
  that none of these are entangled with
  lines outside the subset.
\item Control structures like conditional jumps and loops are not
  easily represented in a circuit diagram.

\end{enumerate}

\section{Review of existing techniques}
Representing quantum algorithms in a way suitable for classical
programmers has been an area of considerable interest. In this
section we discuss some of the techniques currently
proposed. Many of these approaches are implemented with a
simulator back end. 

The different approaches fall into two categories, those that
rely on a graphical user interface (GUI), and those that rely on
text-based programming. 

All the graphical representations  

A slew of graphical applications built upon the
idea of circuit diagrams have been
proposed\cite{quantiki}. Most of these allow 
the user to construct a quantum circuit by dragging and dropping
out-of-the-box components on a panel. The simulators of this
genre mainly differ from one another in terms of the following
points.

\begin{enumerate}

\item The number of
available components, 

\item The ease with which a new component can be
created by the user,
\item the way multi-qubit registers are
visualized ({\em e.g.}, \cite{shary} uses Bloch's sphere to show
  the individual qubits in an unentangled state;\cite{jquantum}
  uses a colour-coded complex plane).
\item The way the measurements are
presented ({\em e.g.}, as colour bars, or as floating
  point numbers or using symbolic expressions). 

\end{enumerate}

Unfortunately we have not seen any GUI based quantum programming
tool that allows the programmer to implement the ``quantum
wrapper'' mentioned earlier. Nor do these tools allows
loops. These two serious drawbacks restrict the uses of these
tools to introductory didactic purposes only.

The second category of quantum programming tools consist of
text-based programming. One conspicuous example is QCL \cite{qcl}
which is a standalone quantum programming language. Other
examples are {\bf Q} proposed by \cite{bettelli} and {\bf LanQ}
by \cite{lanq}. 
A quantum functional programming language QFC has been suggested
by \cite{selinger04}. Excellent (though somewhat antiquated) surveys are provided in \cite{gay,rudiger}.
 
A third approach suggested in \cite{noson} consists of buiding a quantum assembly
language that can be emebedded in a classical program. Some
typical quantum assembler directives could be

\begin{Verbatim}[commandchars=\\\%\#]

INITIALIZE X 1
U TENSOR H I2
APPLY U X

\end{Verbatim}

The quantum (assembly) languages belonging to the second and
 third categories above, 
 while superficially akin to their classical brethren,
nevertheless deprives the programmer of the intuitive feel of a quantum
circuit. The state of a classical program is typically stored in a
 collection of variables, and the programmer processes the
 different variables differently.  This, unfortunately, is not possible in presence of
entanglement, because we cannot store entangled qubits separately. 
In the circuit of {\rm Fig \ref{im:ckt3.eps}}, for example, we cannot really store the
outcome of the first stage into 3 qubits and feed the top two
qubits to $C!$

Our QuECT approach detailed below aims to rectify these flaws.

\section{The QuECT paradigm}
The QuECT paradigm is a hybrid approach where we embed a quantum
circuit diagram in a classical ``host'' program to get the best
of both worlds. 
We use the classical programming constructs for the
classical part, and seamlessly integrate it with circuit diagrams
for the quantum parts. Before we delve into the details let us
take a look at a simple example. 

\subsection{An example}
Consider the quantum circuit shown in {\rm Fig \ref{im:ckt4.eps}}.

\chhobi{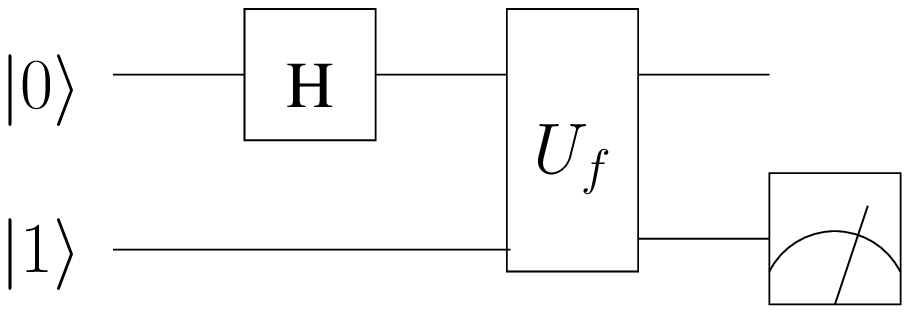}
       {0.4}
       {A circuit to
  demonstrate QuECT}

Here $U_f$ is the quantum wrapper around the classical
function $f:\{0,1\}\to\{0,1\}$ given by $f(x)=1-x.$

The QuECT version of this quantum algorithm is given below using
Java as the classical ``host'' language. The embedded circuit is
shown in bold.

\begin{Verbatim}[numbers=left,commandchars=\\\%\#]
public class QTester {

  public static void main(String args[]) {
    QMachine qm = new QMachine(2);
    H = MatUnitary.HADAMARD;
    Classical c = new Classical(1,1) {//domain dim=1=range dim,
         public f(int x) {return 1-x;}
    };
    Uf = new WrapUnitary(c);
    \ttbf%QBEGIN(qm)#
    \ttbf%|0>-[H]--|Uf|---#
    \ttbf%|1>------|Uf|--->#
    \ttbf%QEND#

    int measVal = qm.getObsDist();
    System.err.println("measured value = "+measVal);
  }
}
\end{Verbatim}

   \begin{description}
   \item[Line 4] A new quantum machine is created to handle a
   2-qubit register. 
   \item[Lines 5] A new Hadamard gate is created. This gate is
   one of the standard gates used in quantum computation.
\item[Lines 6--8] The classical function $f$ is defined.
\item[Line 9] A quantum wrapper is put around this $f.$
\item[Line 10--13] The circuit diagram is embedded as an ASCII 
art between the 

\begin{Verbatim}[commandchars=\\\%\#]
QBEGIN(qm)
  ...
QEND
\end{Verbatim}

delimiters. The name of the target quantum machine ({\tt qm},
  here) is provided as an argument. The syntax of the ASCII  art
  (self-evident in this example) will be explained shortly.
 The `$>$' symbol at the end of line 12 marks that qubit for
  measurement. 
\item[Line 15] The method {\tt getObs()} extracts
the measured value.
\end{description}

\subsection{QuECT syntax}
A
quantum algorithm is embedded in a QuECT program as one or more
chunks of the form
the 

\begin{Verbatim}[commandchars=\\\%\#]
QBEGIN(%\em%name of quantum machine##)
...
QEND
\end{Verbatim}
 
Just
like classical statements, a chunk can be inserted anywhere inside a program ({\em e.g.}, in the body of the
{\tt for}-loop). The syntactic elements of QuECT come in two
flavours---those for use inside a quantum chunk, and those that
are used outside. We start our description with the first
category. 

The syntax for use inside a chunk is
designed to 
mimic a quantum circuit diagram as closely as possible with
ASCII art. Certain features are added to avoid ambiguity during
parsing. A walk-through follows.

\begin{itemize}

\item Each qubit line is shown with a sequence of dashes (the length is
immaterial). Multiple lines can be abbreviated as shown in
{\rm Fig \ref{im:ckt5.eps}}.

\chhobi{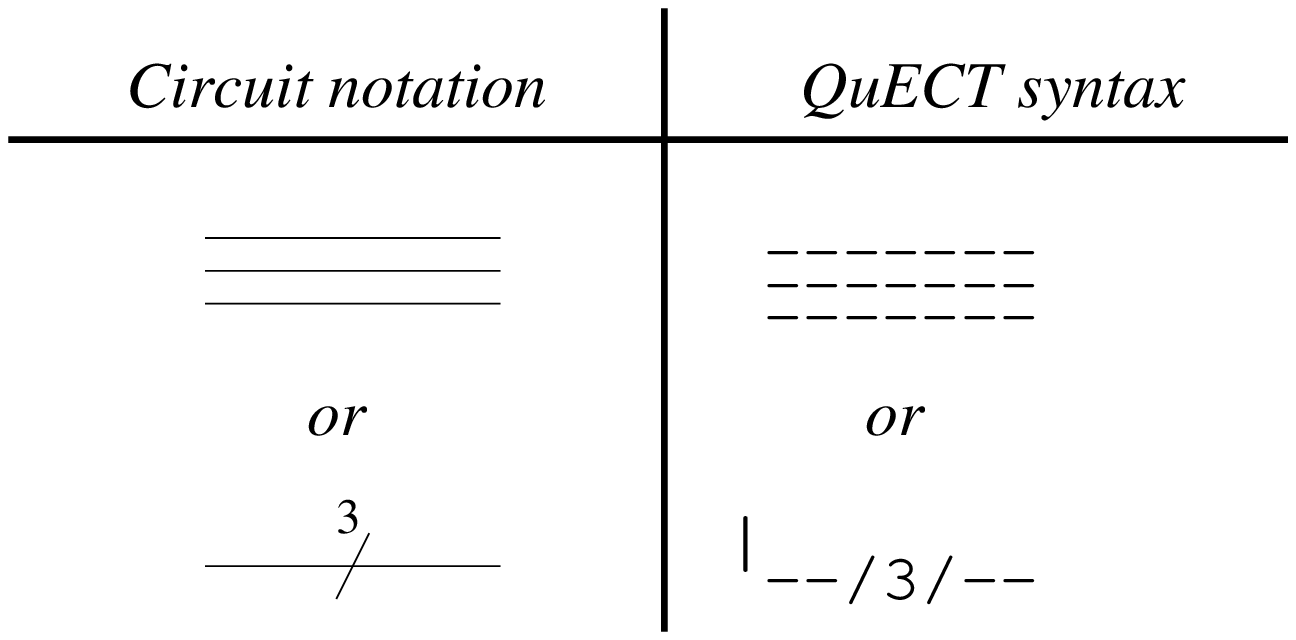}
       {0.4}
       {QuECT syntax for
  single and multi-qubit lines}

We can also use 

\begin{Verbatim}[commandchars=\\\%\#]

---/n/----

\end{Verbatim}

where {\tt n} is some integer variable defined in the classical
part of the QuECT program.

\item Initialization is done as in {\rm Fig \ref{im:ckt6.eps}}.

\chhobi{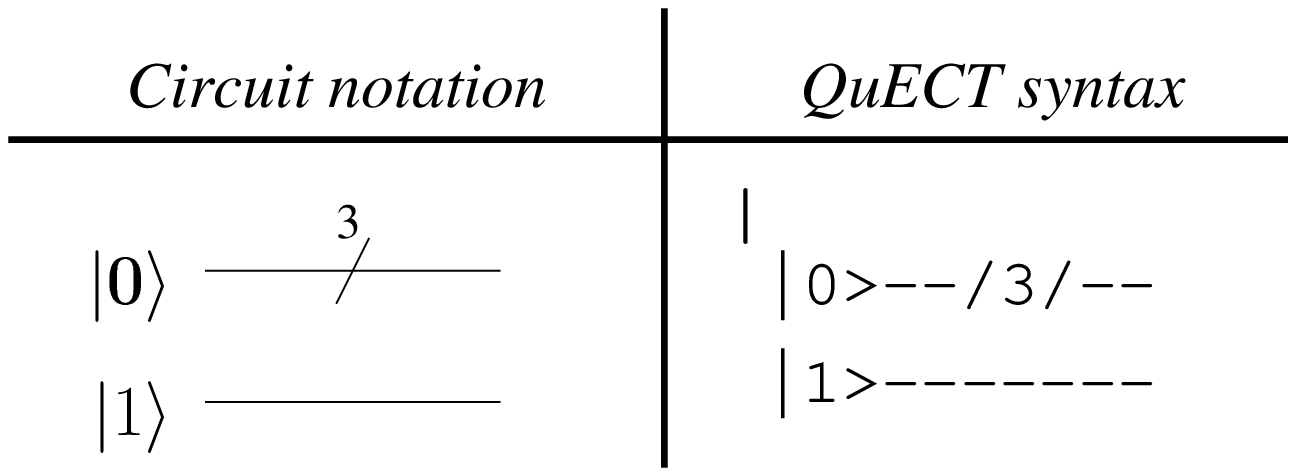}
       {0.4}
       {QuECT syntax
  for initialization}

\item 
The quantum gates spanning only a single line 
are shown inside square brackets ({\rm Fig \ref{im:ckt6.eps}}).

\chhobi{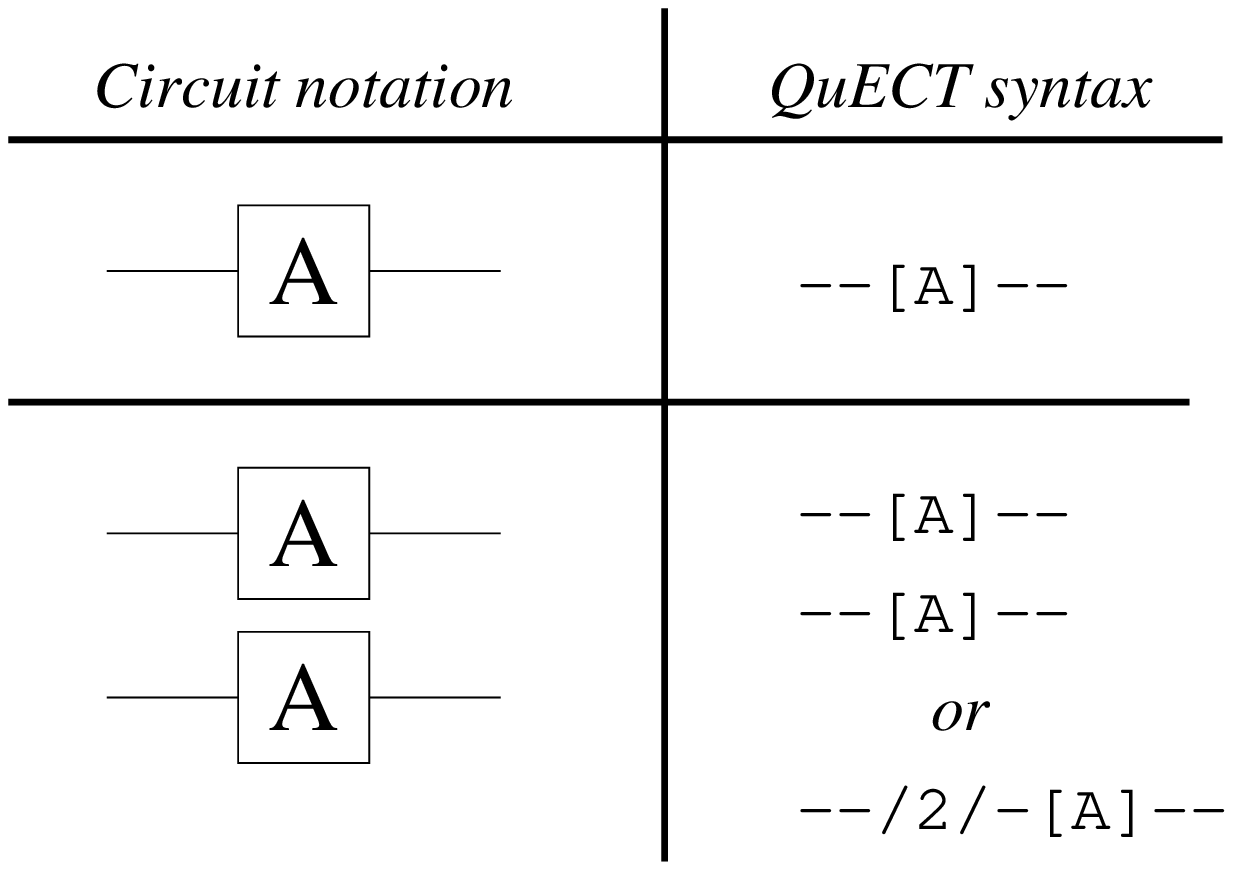}
       {0.4}
       {QuECT syntax for
  single line gates}

\item 
Gates spanning multiple lines are delimited by vertical bars ({\rm Fig \ref{im:ckt7.eps}}). 

\chhobi{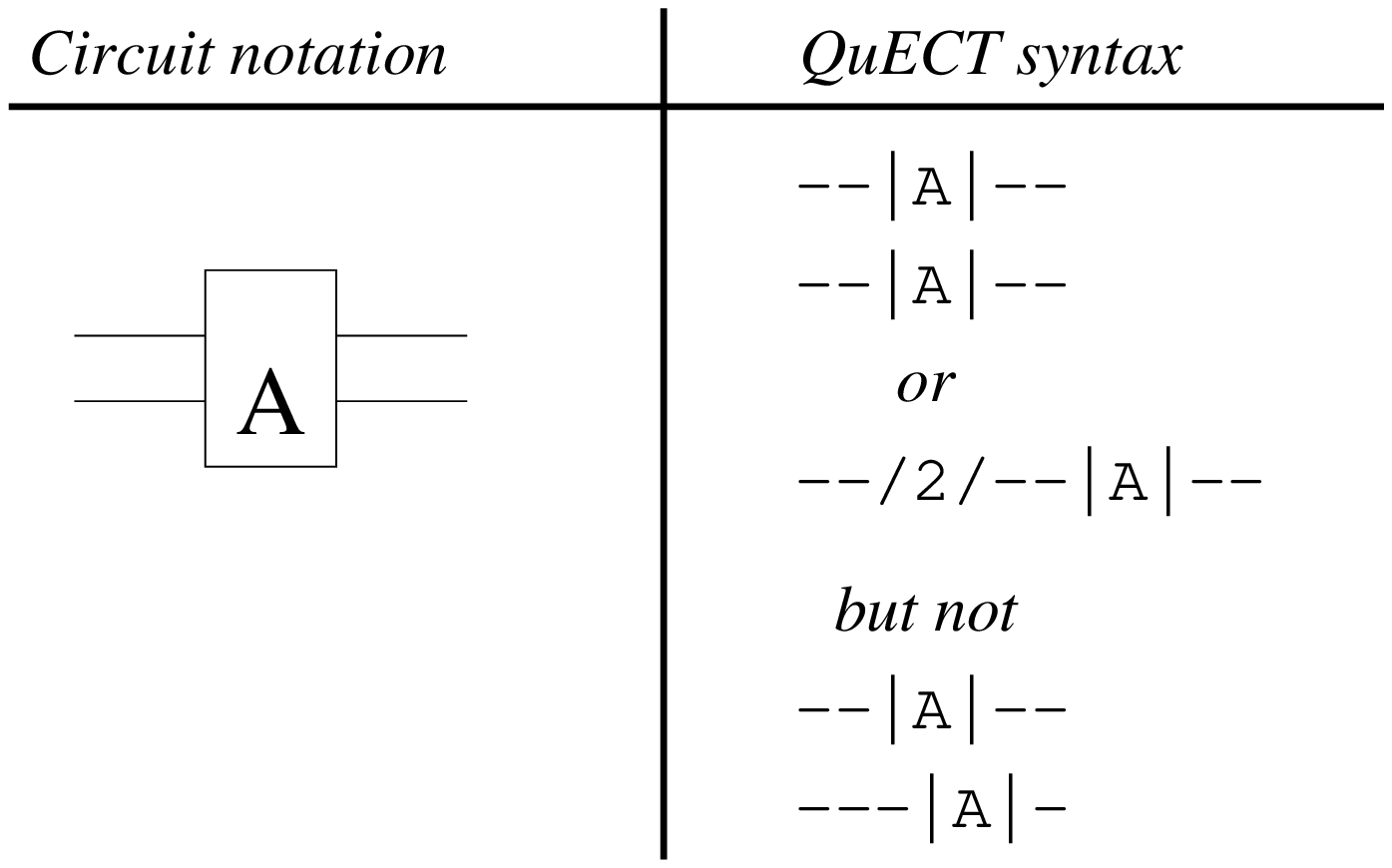}
       {0.4}
       {QuECT syntax for
  multiline gates}

They must be vertically aligned (otherwise a syntax error will be generated).

\item 
Care must be exercised to distinguish the two
situations depicted in {\rm Fig \ref{im:ckt8.eps}}.

\chhobi{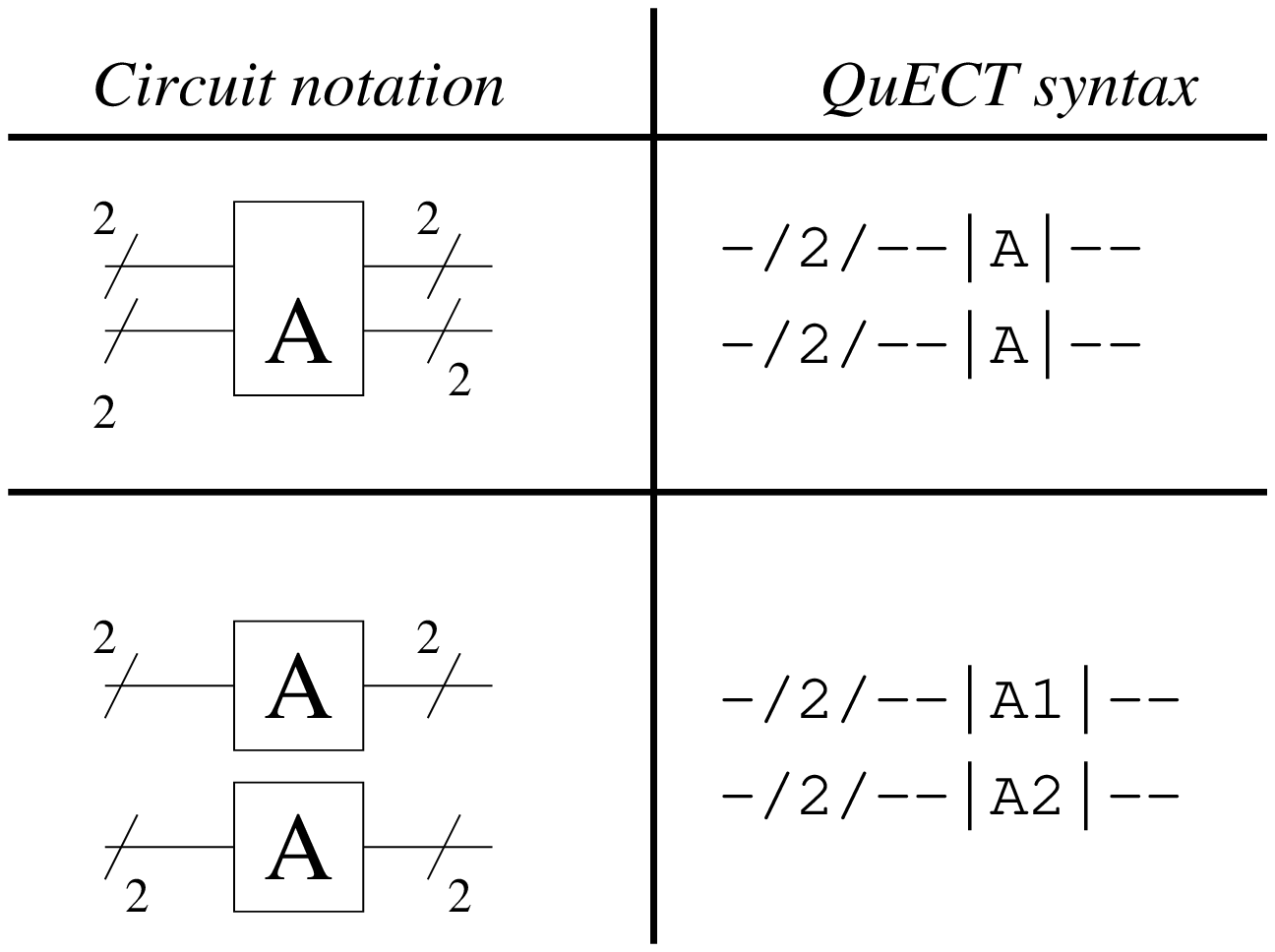}
       {0.4}
       {Two similar but
  different circuits}

Notice the need of two different identifiers {\tt A1}
 and {\tt A2} in the latter case. We could use any other
identifiers also, as long as they are both associated with the
same gate in the classical part of the algorithm.

\item Sometime we need to feed two or more nonadjacent qubits
  into a gate. This often messes up the circuit diagram. Some
  simulators employ a (pseudo-)gate called {\em swap} to bring
  the qubit lines to adjacent positions before feeding them into
  the gate. But  this is easy to achieve in QuECT as
  in {\rm Fig \ref{im:cktadj.eps}}.

\chhobi{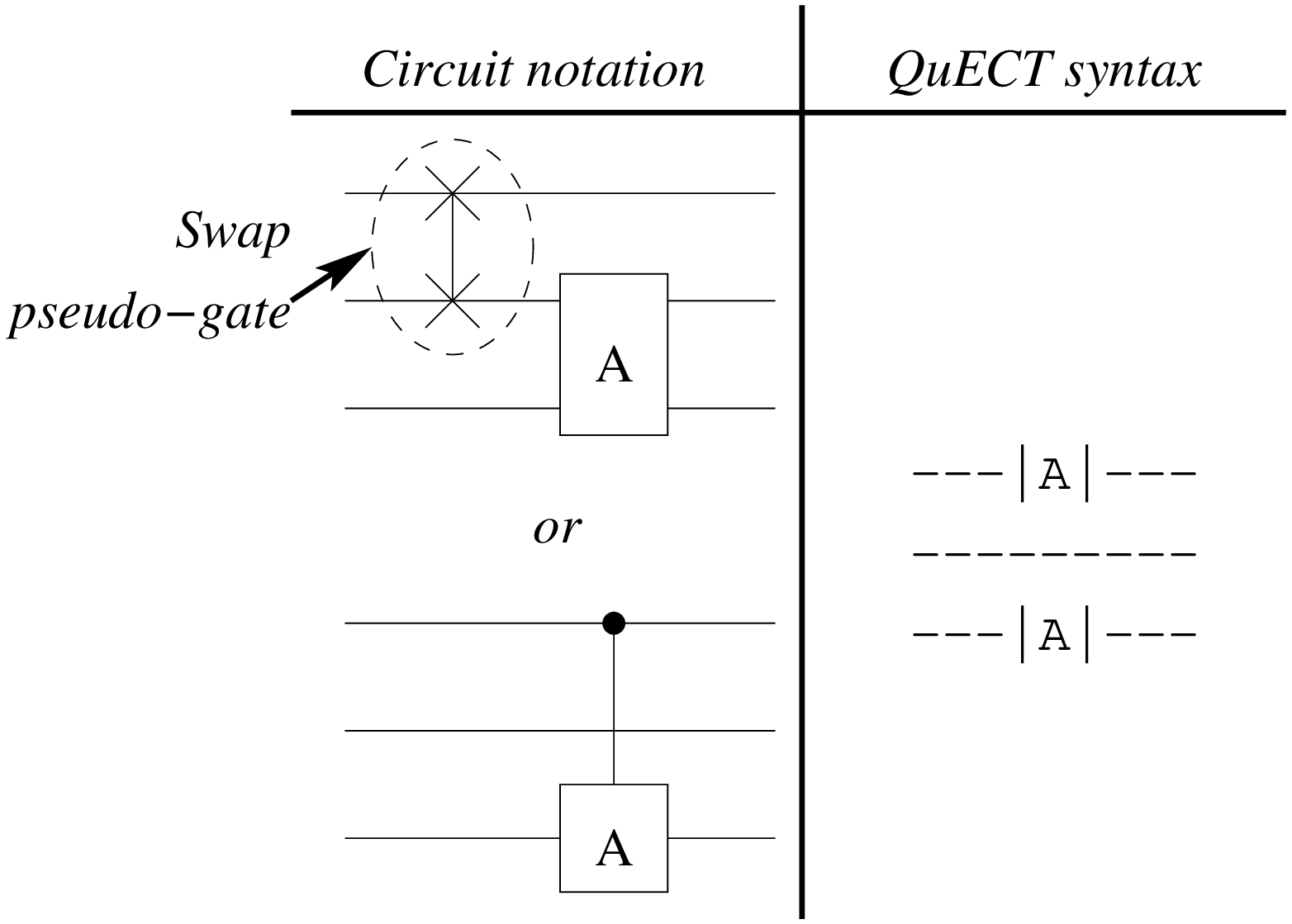}
       {0.4}
       {QuECT syntax
  for gates spanning nonadjacent lines}

Here the gate $A$ takes the two extreme qubits as input,
letting the middle qubit ``pass through''.

\item 
 Swap (pseudo-) gates are somewhat like {\tt goto}-statements,
 and should be generally avoided in a 
circuit diagram, as they reduce readability of the diagram. A
 judicious layout of the lines can avoid the need of swap gates
 in many situations. Also the ``pass through'' syntax of QuECT as
 discussed above reduces the need of swapping lines. But still
 there may be situations where swapping is needed. QuECT provides
 the syntax shown in {\rm Fig \ref{im:cktswap.eps}} for such rare occasions.  

\chhobi{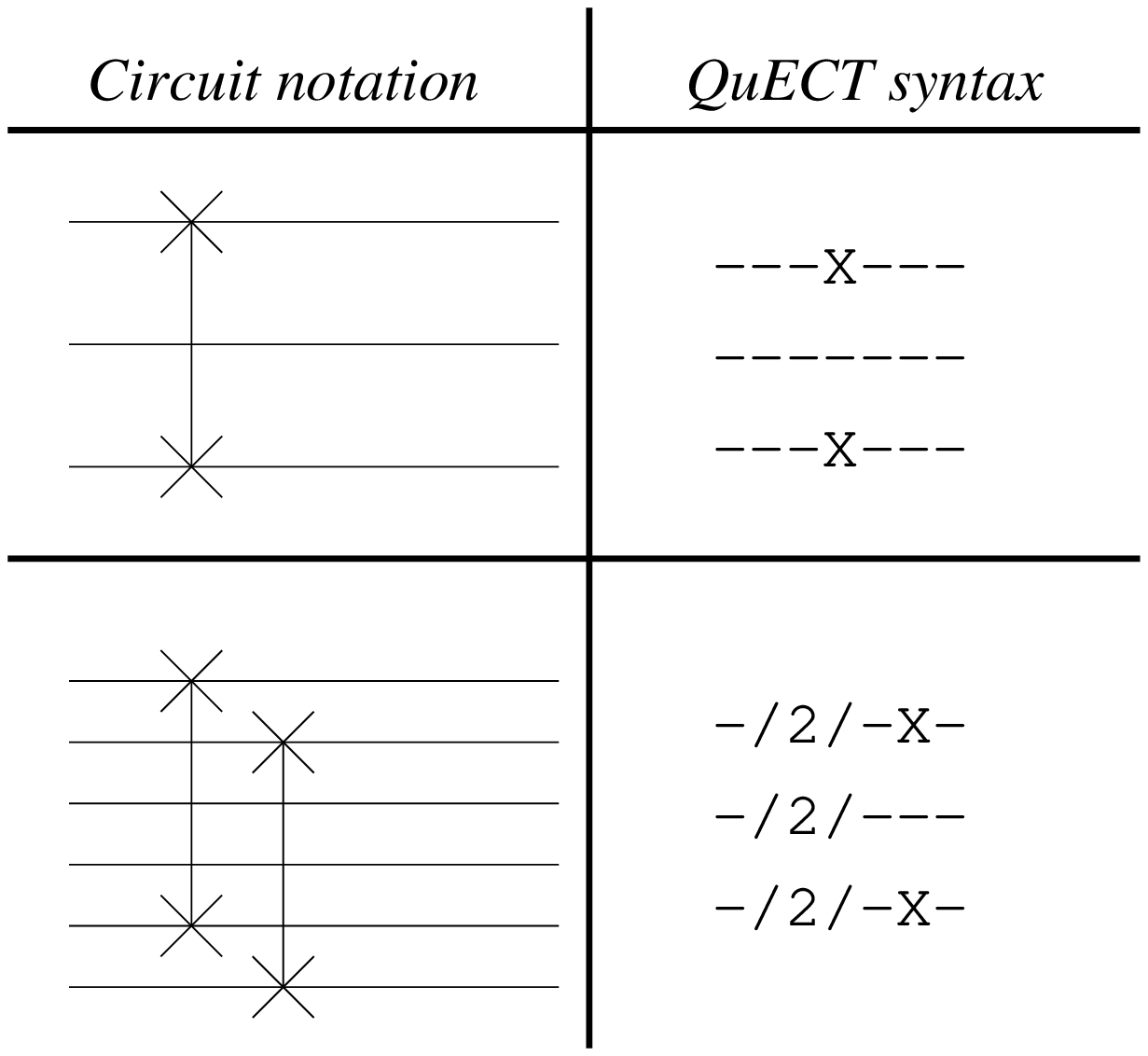}
       {0.4}
       {QuECT syntax
  for swapping lines}

The two {\tt X}'s must be vertically aligned. Also each column
allows either zero or exactly 2 {\tt X}'s. Attempt to swap lines
  with different repeat counts generates syntax error.

\item All measurements are done at the very end. A qubit to be
  measured is marked with a `$>$' at the end of the line ({\rm Fig \ref{im:ckt9.eps}}).

\chhobi{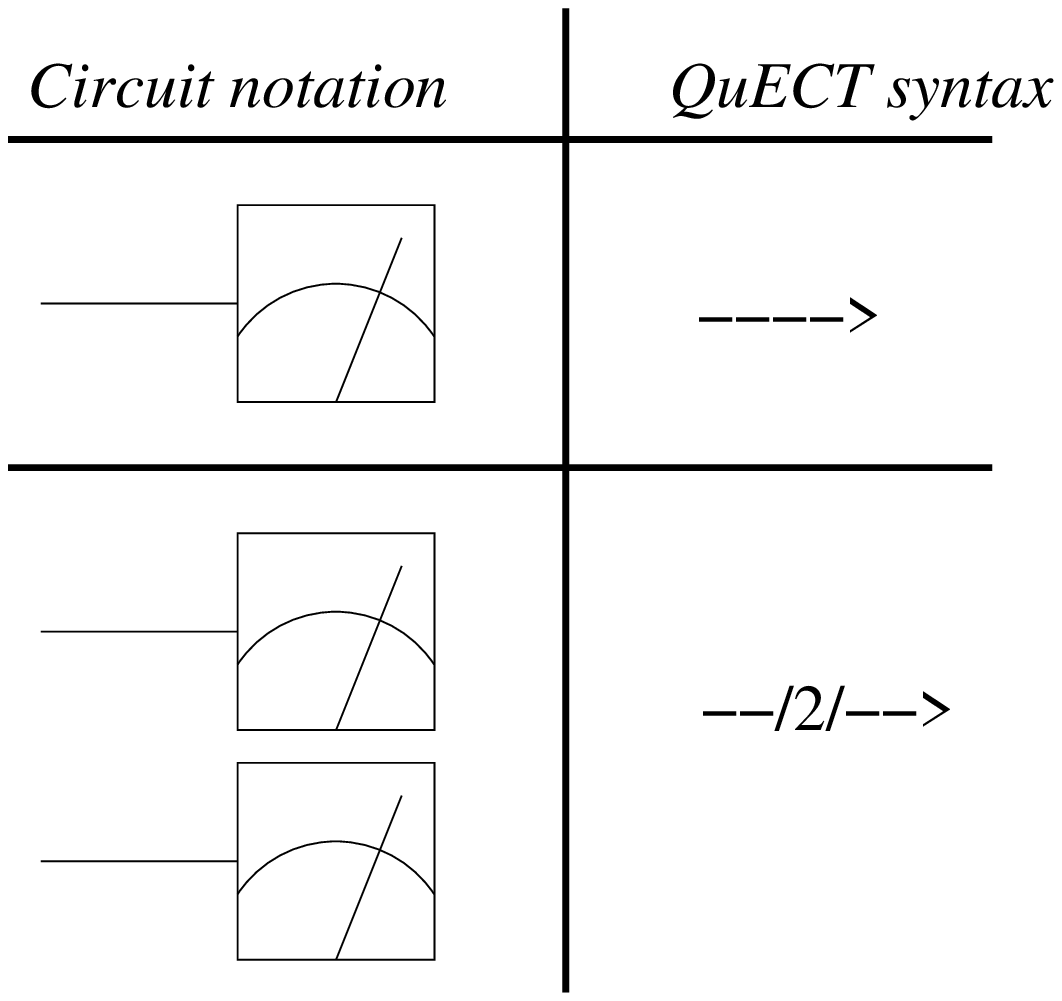}
       {0.4}
       {QuECT syntax for measurement}

\end{itemize}

Next we come to the QuECT syntax for use outside the quantum
chunks.
The actual quantum computation is encapsulated inside a class
called {\tt QMachine}. In a real quantum computer this class will
be responsible for interfacing the underlying quantum
hardware. Alternatively, this class may just run a
simulator. Multiple instances of {\tt QMachine} may exist in
parallel. We interact with the quantum
machine at the three levels discussed below.

\begin{description} 
\item[Construction:\ ]
  
We pass the number of qubits via its constructor.

\item[Creating quantum gates:\ ]
  
The quantum gates can be created in two ways. Either they are
out-of-the-box standard gates ({\em e.g.}, Hadamard), or they are classical functions in
a quantum wrapper. The class {\tt MatUnitary} deals with the
out-of-the-box gates. The {\tt WrapUnitary} class provides a
quantum wrapper for a classical function 
$f:\{0,1\}^m\to\{0,1\}^n.$ Such a function is specified
by $m,n$ and $f,$ which are encapsulated inside the
class {\tt Classical}. Lines 6--8 of the example QuECT code show
an example with $m=n=1$ and $f(x)=1-x.$

\item[Measurements:\ ]
  
All measurements are read back into the classical part via the two methods
{\tt getObsDist()} and {\tt getObs()}. The first method returns the probability
distribution as an array of length $2^k$ where $k$ is
the number of qubits being measured. The second method returns an
actual output. A simulator may use
randomization to generate one output from the output  distribution. In
an actual quantum computer, only the second method will be
available. Explicitly computing the probability distribution of
the measurement is a luxury that we can afford only in a simulator. 

\end{description}
\section{Some standard algorithms}
In this section we show the implementation of some well known
quantum algorithms using QuECT. The aim is not to acquaint the
readers with the details of the algorithms, rather to demonstrate
different applications of the new paradigm. Interested readers
will find a very readable account of the algorithms in \cite{noson}. 

\subsection{Deutsch algorithm} 
The simplest possible quantum algorithm\cite{noson} is the Deutsch algorithm
which checks if a given classical ``blackbox'' function 
$$
f:\{0,1\}\to\{0,1\}
$$
is 1-1 or not.

\begin{Verbatim}[numbers=left,commandchars=\\\%\#]
QMachine qm = new QMachine(2,1);

Classical c = new Classical(1,1) {
   public f(int x) {return 1-x;}
};
Unitary Uf = new WrapUnitary(c));

\ttbf%QBEGIN(qm)#
\ttbf%|0>--[H]--|Uf|---[H]-->#
\ttbf%|1>--[H]--|Uf|---------#
\ttbf%QEND#

if(qm.getObs()==1)
  System.out.println("1-1");
else
  System.out.println("not 1-1");
\end{Verbatim}

\subsection{Deutsch Jozsa algorithm}
This is a multidimensional generalization of Deutsch algorithm
from the last section\cite{deujoz}. Here we start with a classical
``blackbox'' function $f:\{0,1\}^n\to\{0,1\},$ which is
known to be either a constant function or a ``balanced'' function
, {\em i.e.}, exactly half of the $2^n$ binary $n$-tuples
are mapped to 0, the other half being mapped to 1.  

The aim of the algorithm is to detect which is the case. 

\begin{Verbatim}[numbers=left,commandchars=\\\%\#]
int n = 4;
QMachine qm = new QMachine(n+1);

Classical c = new Classical(n,1) {
   public f(int x) {return x & 1;} //A sample balanced function
};
Uf = new WrapUnitary(c);

\ttbf%QBEGIN(qm)#
\ttbf%|0>--/n/--[H]--|Uf|---[H]-->#
\ttbf%|1>-------[H]--|Uf|---------#
\ttbf%QEND#
\end{Verbatim}

\subsection{Simon's periodicity algorithm}
Here we are given a classical ``blackbox''
function $f:\{0,1\}^n\to\{0,1\}^n,$ which is known to be
periodic, {\em i.e.},  
$$
\exists\bb\in\{0,1\}^n\text{ such that } \forall \bx\in
\{0,1\}^nf(\bx\oplus\bb) = f(\bx).
$$
We are told that such a $\bb$ exists, but we do not know
what $\bb$ actually is. Simon's algorithm\cite{simon} is a way to find
this $\bb.$ The algorithm starts with some quantum
computation followed by a classical linear equation solver. We
present only the quantum part here.

\begin{Verbatim}[numbers=left,commandchars=\\\%\#]
int n = 5, orthog[100];
QMachine qm = new QMachine(2*n);

Classical c = new Classical(n,n) {
   public f(int x) {//f is defined here}
};
Uf = new WrapUnitary(c);
for(int i=0;i<100;i++) {
  \ttbf% QBEGIN(qm)#
  \ttbf% |0>--/n/--[H]--|Uf|---[H]-->#
  \ttbf% |0>--/n/-------|Uf|---------#
  \ttbf% QEND#
  orthog[i]=qm.getObs();  
}
\end{Verbatim}

 Notice how we have put a quantum chunk inside a {\tt for}-loop.
\subsection{Grover's search algorithm}
Grover's search algorithm\cite{grover} is for searching a linear
list of size $2^n.$

\begin{Verbatim}[numbers=left,commandchars=\\\%\#]
int n = 6;
QMachine qm = new QMachine(n+1);

Classical c = new Classical(1,1) {
   public f(int x) {return (x = 7? 1 : 0);} //We are searching
                                            //for 7 in {0,...,63}.
};

Uf = new WrapUnitary(c);

\ttbf% QBEGIN(qm)#
\ttbf% --/n/--[H]-#
\ttbf% -----------#
\ttbf% QEND#

for(int i=0;i<100;i++) {
  \ttbf% QBEGIN(qm)#
  \ttbf% --/n/--|Uf|--[IM]--#
  \ttbf% --[H]--|Uf|--#
  \ttbf% QEND#
}

\ttbf% QBEGIN(qm)#
\ttbf% --/n/-->#
\ttbf% -------#
\ttbf% QEND#
\end{Verbatim}

Here we have used multiple quantum chunks. The quantum
              gate $IM$ performs an operation called
{\em inversion around mean}, whose details need not concern us here. 
\subsection{Shor's algorithm and QFT}
This algorithm factorizes a given integer in polynomial time\cite{shor}. 
Its quantum part is structurally quite similar to the
algorithms already discussed, except for a Quantum Fourier Transform (QFT) block. 
The definition of $QFT$ is recursive. If we
denote the $n$-qubit QFT gate by $Q_n$ then it is
defined recursively in terms of $Q_{n-1}.$ A typical step
(for $n=4)$ is shown in {\rm Fig \ref{im:qft.eps}}.

\chhobi{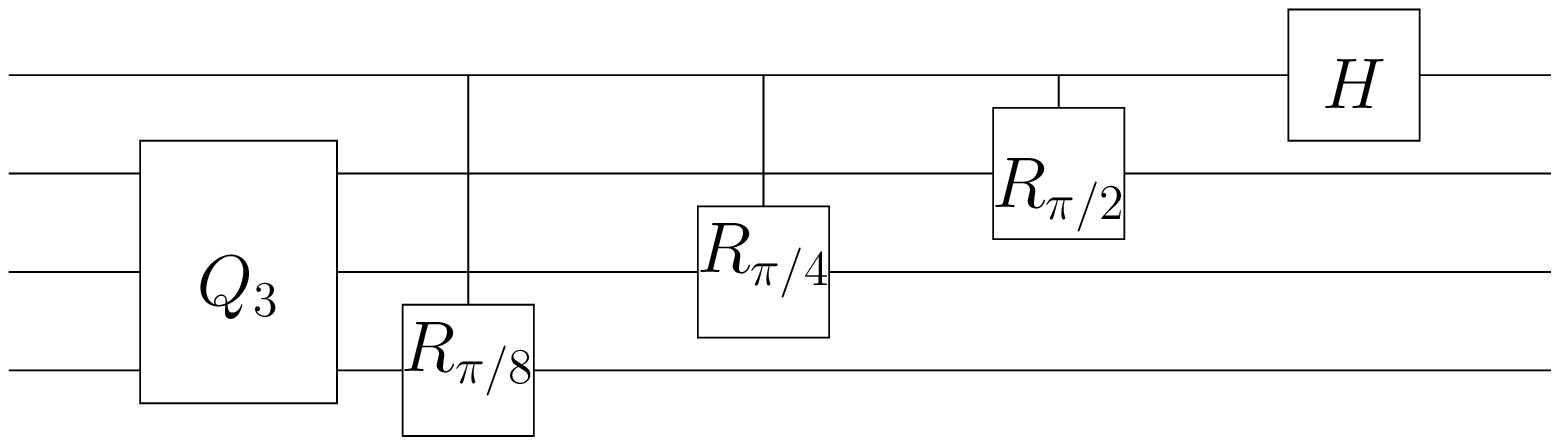}
       {0.4}
       {Definition
  of $Q_4$ in terms of $Q_3$ }

Two points set this circuit aside from the ones already
discussed earlier: recursion and the $R$ gate which has a
parameter (shown in the subscript). We shall not go into the
details of the $R$ gate. We shall just treat it as a
blackbox and show the implementation of the circuit in QuECT.

We notice that the basic building
blocks are 

\chhobi{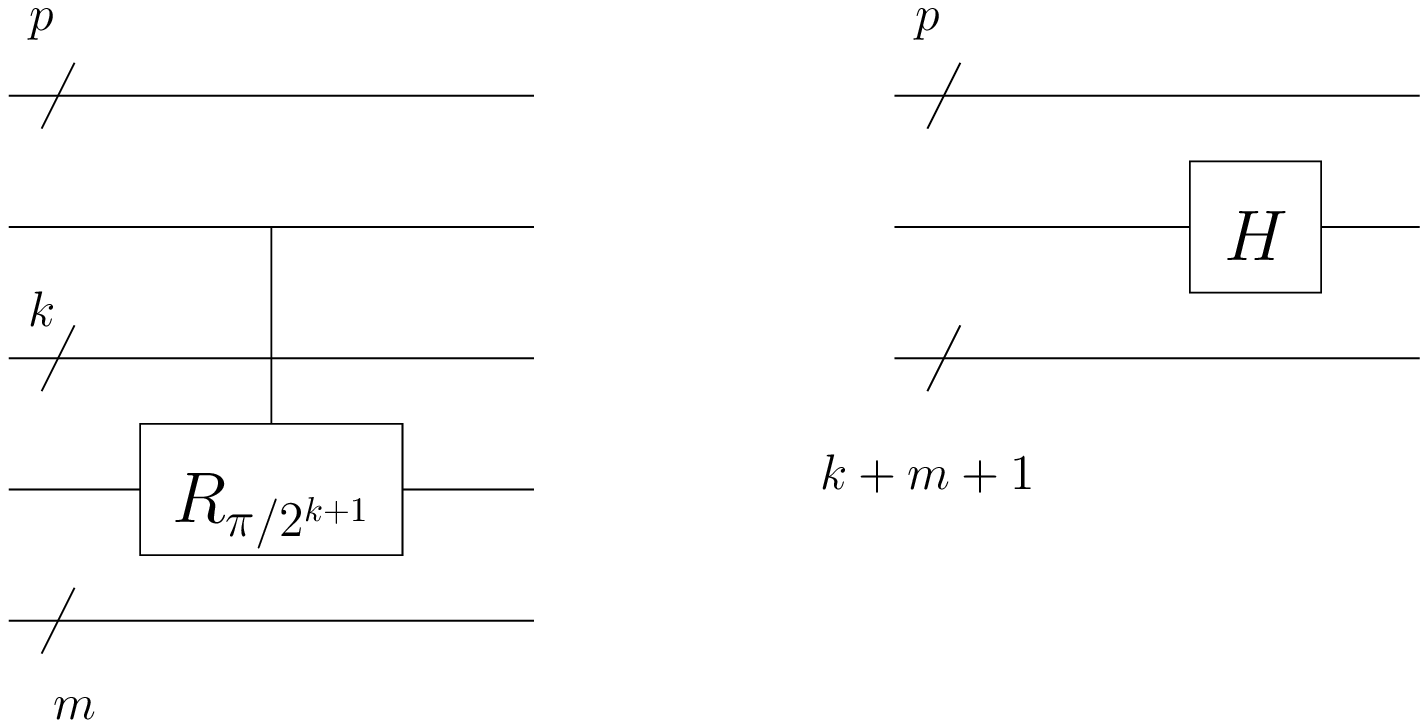}
       {0.4}
       {}

Let us construct $Q_n$ using these. There are $n$
recursive steps. In the $r$-th step $(0\leq r\leq n-1)$we define $Q_{r+1}$ in
terms  of $Q_r.$ This requires $r$ applications of
the first block, followed by one application of the second
block. It is not difficult to see that this is achieved by the
following hybrid code.

\begin{Verbatim}[numbers=left,commandchars=\\\%\#]
for(int r=0;r<n;r++) {
  int p = n-(r+1);
  for(int k=1;k<r;k++) {
    int m = r-(k+1);
    Unitary R = new Rgate(Math.PI/(2 << k));
    QBEGIN(qm)
    --/p/-------
    ------------
    --/k/-------
    ------[R]---
    --/m/-------
    QEND
  }

  int q = r-1;
  QBEGIN(qm)
  --/p/-------
  -------[H]--
  --/q/-------
  QEND
}
\end{Verbatim}

 Notice how we are performing both classical as well as quantum
 computation in the same pass of a {\tt for}-loop.

\section{Implementation}
In this section we briefly touch upon the issue of compiling a
QuECT program. As QuECT is not a programming language, but a
paradigm, it does not make strict sense to compile a QuECT
program. The details will depend on the classical host
language. But certain host-independent suggestions can be given,
and this is our aim here.

 Typically the compilation of any computer program
proceeds in two stages: the front end reduces the program to some
abstract intermediate representation, while the back end
converts this representation to hardware specific code. 

For a QuECT program we suggest that the back end should produce
code in the host language with the {\tt QMachine} class
encapsulating all the quantum hardware details. 
The following
describes one approach for the front end.

The information contained inside a single quantum chunk
can be represented as
follows. 

\begin{enumerate}

\item The number of lines. Here a multi-qubit line counts as
  a single line.
\item A list of repeat counts, one for each line.
\item A list of {\em Stage}s, where each {\em Stage} consists of
a list of gates (including swap pseudo-gates), each with its list of lines.

\item A list of lines to be measured.

\end{enumerate}

An example will make this clear. 

Suppose that we have

\begin{Verbatim}[numbers=left,commandchars=\\\%\#]
QBEGIN
-/n/--|A|----|Uf|---
--------[B]--|Uf|--->
------|A|----------
QEND
\end{Verbatim}

Here we have 3 lines, the repeat counts are $n,1$ and 1.
There are two stages. The first stage consists of two
  gates, $A$ (spanning lines 0 and 2)
                    and $B$ (spanning line 2). 
Measurement is done only on line 1. 

It is now easy to embed this information to a suitable method in
                    the {\tt QMachine} class, producing an
                    ordinary Java program.
The advantage of producing the output as a high level host
                    program
is that we can borrow the symbol table of the host language,
and use the identifiers declared in the classical part.  

\section{Conclusion}
In this paper we have proposed a new paradigm called Quantum
Embedded Circuit Technique (QuECT) for programming
a quantum computer. Our approach seeks to combine the advantage
of classical programming constructs as well as the visual
benefits of a quantum circuit diagram. The paradigm can be easily
implemented in any high level language, preferably with
object-oriented facilities. At present this can be
used as a powerful and versatile front end to simulators. The same
approach can be easily applied to harness the power of the real
quantum computers when they are built.

\bibliographystyle{plain}
\bibliography{test}

\end{document}